\documentclass[journal=jacsat,manuscript=article,layout=twocolumn]{achemso}
\usepackage[version=3]{mhchem} % Formula subscripts using \ce{}
\usepackage{xcolor}
\usepackage{siunitx}

\usepackage[normalem]{ulem}
\usepackage{soul}
\usepackage{cancel}
\usepackage{easyReview}
\usepackage{cuted}

\usepackage{lipsum}

\usepackage[frozencache,cachedir=.]{minted}
\author{Albert K. Engstfeld}
\affiliation[Ulm University]
{Institute of Electrochemistry, Ulm University, D-89069 Ulm, Germany}
\email{albert.engstfeld@uni-ulm.de}
\author{Johannes M. Hermann}
\affiliation[Ulm University]
{Institute of Electrochemistry, Ulm University, D-89069 Ulm, Germany}
\author{Nicolas G. Hörmann}
\affiliation{Fritz-Haber-Institut der Max-Planck-Gesellschaft, Faradayweg 4-6, 14195 Berlin, Germany}
\author{Julian Rüth}
\affiliation[Ulm University]
{Julian Rüth GmbH, Weinbergstr. 48, D-97261 Güntersleben}
\title[echemdb toolkit]
  {echemdb Toolkit - a Lightweight Approach to Getting Data Ready for Data Management Solutions.}

\abbreviations{FAIR, Metadata}

\begin{document}

%\begin{tocentry}

%Some journals require a graphical entry for the Table of Contents.
%This should be laid out ``print-ready'' so that the sizing of the

%Inside the \texttt{tocentry} environment, the font used is Helvetica
%8\,pt, as required by \emph{Journal of the American Chemical

%The surrounding frame is 9\,cm by 3.5\,cm, which is the maximum
%permitted for  \emph{Journal of the American Chemical Society}
%graphical table of content entries. The box will not resize if the
%content is too big: instead it will overflow the edge of the box.

%This box and the associated title will always be printed on a
%separate page at the end of the document.

%\end{tocentry}

\begin{abstract}
According to the FAIR (findability, accessibility, interoperability, and reusability) principles, scientific data should always be stored with machine-readable descriptive metadata. Existing solutions to store data with metadata, such as electronic lab notebooks (ELN), are often very domain-specific and not sufficiently generic for arbitrary experimental or computational results.

In this work, we present open-source echemdb toolkit for creating and handling data and metadata. The toolkit is running entirely on the file system level using a file-based approach, which facilitates integration with other tools in a FAIR data life cycle and means that no complicated server setup is required. This also makes the toolkit more accessible to the average researcher since no understanding of more sophisticated database technologies is required.

We showcase several aspects and applications of the toolkit: automatic annotation of raw research data with human- and machine-readable metadata, data conversion into standardised frictionless Data Packages, and an API for exploring the data. We also illustrate the web frameworks to illustrate the data using example data from research into energy conversion and storage. 
\end{abstract}

\newpage


%%%%%%%%%%%%%%%%%%%%%%%%%%%%%%%%%%%%%%%%%%%%%%%%%%%%%%%%%%%%%%%%%
%%%%%%%%%%%%%%%%%%% --- Introduction --- %%%%%%%%%%%%%%%%%%%%%%%%
%%%%%%%%%%%%%%%%%%%%%%%%%%%%%%%%%%%%%%%%%%%%%%%%%%%%%%%%%%%%%%%%%
\section{Introduction}

Handling scientific data according to the FAIR principles\cite{WILKINSON2016,BAHIM2020} and covering all aspects of the data life cycle can be an arduous task when no research data management (RDM) infrastructure is available. 
One of the most crucial aspects is that the data is stored in a machine-readable format, which includes additional information describing the object of investigation.
Electronic lab notebooks (ELNs) are suggested to simplify storing the data accordingly. 
However, even in a single institution, data often exists in different formats, ranging from single values (observable at a specific time) to time series data (change of an observable with time), images (such as microscope images), or surveys. Each data set may have different requirements regarding storage, evaluation, and description. Furthermore, the investigated systems are often highly inhomogeneous, i.e., standards for data storage might not exist, and if they exist, they are often very specific to a certain system or domain. Thus, the available ELN solutions are often not suitable.
Finally, choosing the right ELN is difficult, especially when domain-specific RDM experts are unavailable. New researchers, e.g., freshly graduated students, usually lack the necessary skills and (domain) knowledge (RDM or programming) to develop a solid and sustainable RDM infrastructure.

In this work, we present a lightweight approach for automatically annotating research data with metadata in a data-interchange format (here YAML) that is both human- and machine-readable. We then bundle data and metadata in a frictionless Data Package, an open standard data container\footnote{see hdf5,\cite{HDF2021} RO-Crate\cite{ROCRATE2022}, BagIt\cite{BAGIT} for other data container formats} that simplifies exploration, evaluation, validation, and definition of workflows.\cite{JEJKAL2022}

The unitpackage API\cite{UNITPACKAGE2024} plays a central role in simplifying the creation of Data Packages and their exploration by extending the frictionless framework.\cite{FRICTIONLESS2024} 
In some cases, data is only available from figures such as in old publications. We briefly present svgdigitizer,\cite{SVGDIGITIZER2023} which extracts data from such figures and writes frictionless Data Packages that can be explored using the unitpackage API.
Examples of further usage of Data Packages created with our approach include defining a domain-specific metadata schema to annotate the data, turning a collection of Data Packages into a simple database on the file system, which can be explored
%and using the data collection in 
interactively in, i.e., Jupyter notebooks. The workflow from storing to exploring data is illustrated for tabular (time-series) data stored as plain-text CSV (comma-separated values) from the research area of electrochemistry, for which the community has not settled on a standard yet.

In total, our approach does not require a complex RDM infrastructure but only fundamental skills in Python programming.
Data stored using our approach will be more easily located in the file system, improve interoperability for further use in more complex workflows\cite{JEJKAL2022} or ELN solutions, and provide easy access to the data should high-level solutions (ELN) become unavailable.

%%%%%%%%%%%%%%%%%%%%%%%%%%%%%%%%%%%%%%%%%%%%%%%%%%%%%%%%%%%%%%%%%
%%%%%%%%%%%%%%%%%%%% --- Results --- %%%%%%%%%%%%%%%%%%%%%%%%%%%%
%%%%%%%%%%%%%%%%%%%%%%%%%%%%%%%%%%%%%%%%%%%%%%%%%%%%%%%%%%%%%%%%%
\section{Results and Discussion}

In the first section, we discuss the fundamental requirements for storing data for our workflow, 
namely file naming conventions and data exchange formats, and present simple ways to annotate data automatically. 
The second section demonstrates the potential of storing data in containers using the unitpackage API. 
In the final section, we show how data following a metadata schema can be used in more complex workflows, allowing for direct comparison between literature data and local data using the same infrastructure.

\subsection{Data Preparation}

\subsubsection{File Naming Conventions}

When storing experimental data, filenames are still commonly used to store metadata, such as the date, time, user, sample name, or system-specific information such as concentrations, starting weight, or temperature. While this approach allows for convenient searching with file system tools, limitations in file names (length and allowed character set) typically lead to heavy usage of acronyms.

The acronyms used are then often only known to the experimentalist who recorded the data.
While adopting a file naming convention\cite{FNC} used by the peers in the laboratory would mitigate this issue, sharing such data beyond the boundaries of the laboratory is hard, and processing such metadata in a robust way automatically is virtually impossible.
For the approach presented in this work, filenames do not really matter. They only must be unique because they will be used as the unique identifier for the underlying data. Nonetheless, including minimum information in the filename can help organise files in the file system.

\begin{figure*}[t] %Add * to span figure over two columns
    \centering
    \begin{tabular}{p{(\textwidth-1cm)/3}|p{(\textwidth-1cm)/3}|p{(\textwidth-1cm)/3}}
    a) YAML 
\begin{minted}{yaml}
name: Max Doe
date: 2024-06-11
sample name: abc123
amount:
  unit: g
  value: 15
description: A great experiment.
\end{minted}&
        b) JSON
\begin{minted}{json}
{
  "name": "Max Doe",
  "date": "2024-06-11",
  "sample name": "abc123",
  "amount": {
    "unit": "g",
    "value": 15
  },
  "description": "A great experiment."
}
\end{minted}&
        c) XML
\begin{minted}{xml}
<name>Max Doe</name>
<date>2024-06-11</date>
<sample name>abc123</sample name>
<amount>
  <unit>g</unit>
  <value>15</value>
</amount>
<description>A great experiment.</description>
\end{minted}
    \end{tabular} 
    \caption{An example set of metadata for data exchange formats, i.e., a) YAML, b) JSON, and c) XML, storing metadata in key-value pairs.}
    \label{fig:metadata_demo}
\end{figure*}

\subsubsection{Metadata Templates}

Instead of trying to squeeze additional descriptive information about the data in the length-limited filename, the header or the footer of the data file itself, a piece of paper, or another unstructured text file, it is far more beneficial to store this information in a data exchange format such as YAML\cite{YAML2001}, JSON\cite{JSON2006}, or XML\cite{XML1998}.

An example set of metadata is shown for all three file formats in Fig.~\ref{fig:metadata_demo}. For simple metadata files, these formats are interchangeable by various converters. From a practical perspective, YAML is presumably the most human-readable and, hence, accessible format. In addition, YAML can contain comments as inline annotations (starting with a \mintinline{yaml}{#}. This is specifically relevant to the creation of instructive templates for other users.

In contrast to the file name, the metadata templates can now contain an almost limitless amount of information in a structured way on the investigated system, the data itself, the operator or technical details on the experimental setup.
Once the metadata structure has matured and the workflow for data acquisition and evaluation is refined (see below), a metadata schema, such as a JSON Schema\cite{JSONSCHEMA2020}, can be derived. The schema can be used to validate the YAML formated metadata while recording data. Furthermore, such schemas help design metadata templates for more complex ELN systems.

\subsubsection{Metadata Acquisition}

To save time, it is desirable to generate metadata whenever data is created instead of writing metadata manually. This can be achieved by tracking file creation events. 
We illustrate the approach 
using the Python watchdog package.\cite{WATCHDOG}
It provides flexibility in the development phase as it can be tested in a Jupyter environment, as illustrated in the supporting information. 
To unambiguously associate the metadata with a specific file, we suggest creating a metadata file with the same name as the recorded file, including the suffix. Additional suffixes must be added to declare the content of the metadata file. For example, upon creating a \mintinline{text}{data.csv}, a \mintinline{text}{data.csv.meta.yaml} is stored.\cite{SUFFIX} 
When a series of measurements is performed, the user only has to change individual key-value pairs in the template file.

\begin{figure*}[t] %Add * to span figure over two columns
    \centering
    \includegraphics[width=0.7\textwidth]{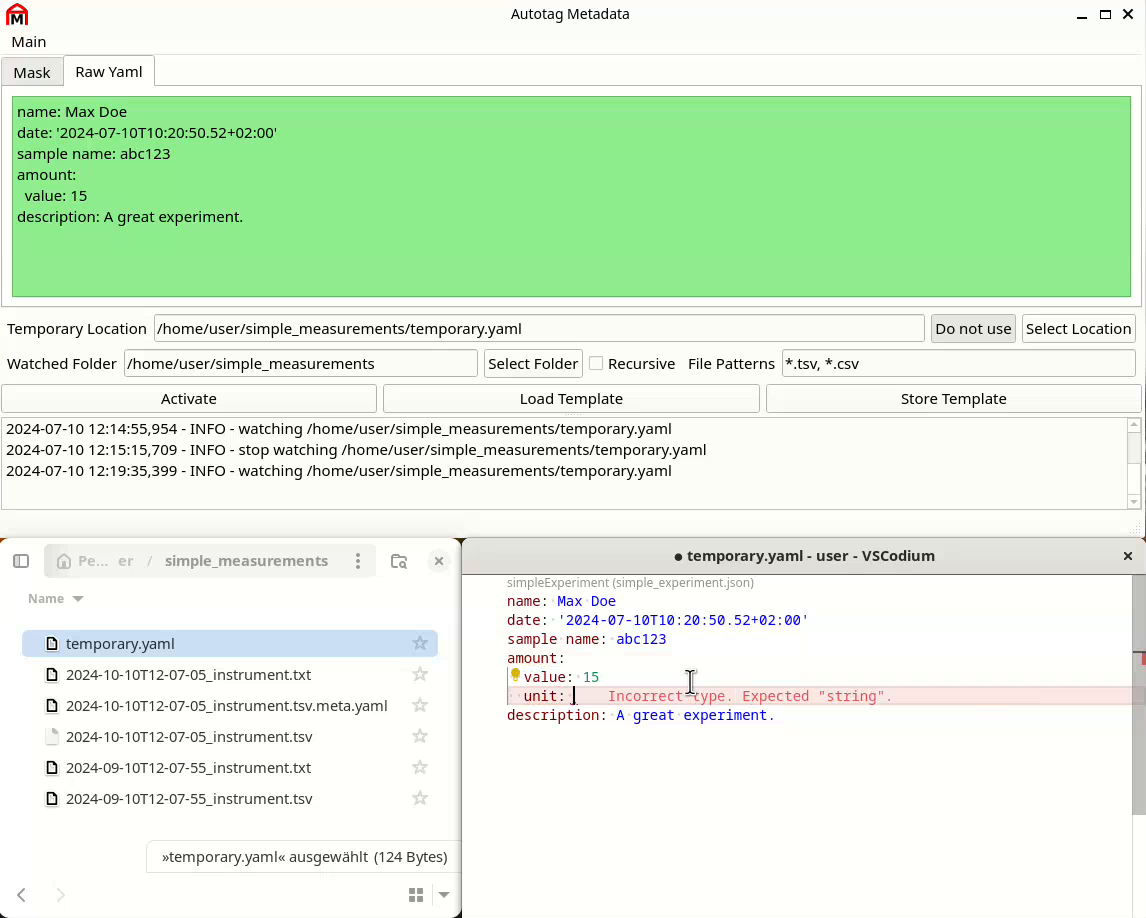}
    \caption{Snapshot of the graphical user interface of the \mintinline{python}{autotag-metadata} program (top),\cite{AUTOTAG2024} which monitors the file system (bottom left) for file creation events and tags files with metadata. The program can be coupled with external editors, such as VSCodium (bottom right), which provides syntax highlighting and can be used for validating the metadata against a schema.}
    \label{fig:autotag}
\end{figure*}

A standalone solution to tag files with metadata employing a graphical user interface (GUI) is desirable for the end-user without programming knowledge. 
For this purpose, we created \mintinline{text}{autotag-metadata} a PyQt-based standalone application.\cite{AUTOTAG2024}
The key components are shown in Fig.~\ref{fig:autotag}. The program watches for file creation events upon which a YAML file containing the metadata is created. The metadata can also be edited in a separate text editor, allowing for advanced syntax highlighting and schema validation, ensuring the highest quality of the acquired metadata. Furthermore, we are exploring an application created in Jupyter, based on ipywidgets (\mintinline{text}{etiquetado-voila}), for use cases requiring more flexibility. These tools are still under development and are currently tailored to the specific needs of our applications.

Note that these tools allow for automatic tagging of files and can be used for any file type, such as text files, PDFs, video clips, or vacation images.

\begin{figure*}[t] %Add * to span figure over two columns
    \centering
    \begin{tabular}{p{(\textwidth-1cm)/2}|p{(\textwidth-1cm)/2}}
    a) CSV 
\begin{minted}{text}
t,U,T
0,1.01,275
1,1.02,275
2,1.05,275
3,0.95,275
4,0.99,275
5,1.01,275
6,0.98,275
7,0.99,275

\end{minted}&
        b) YAML Metadata
\begin{minted}{yaml}
name: Max Doe
date: 2024-06-11
sample name: abc123
amount:
  unit: g
  value: 15
description: A great experiment.
\end{minted}
    \end{tabular} 
    \caption{Example content of files for a) time series data stored as a CSV and b) metadata stored as YAML. The latter describes the structure of the CSV.}
    \label{fig:csv_metadata}
\end{figure*}

\subsubsection{Tabular (Time Series) Data} 

As mentioned above, data can be very different (single values, tabular data, images, etc). Here, we focus on tabular (time series) data that are preferably stored as CSV as shown in Fig.~\ref{fig:csv_metadata}a. A key issue is that the field names (column names) are often not sufficient or incomplete (missing units) to understand the meaning of the values of each field. Also, from a certain dimension, such as the time $t$, we can not infer whether it is a relative or absolute time or whether a temperature $T$ is measured in Kelvin or Celsius.
Such information can also be stored in the YAML template, as illustrated in Fig.~\ref{fig:csv_metadata}b.
The structure of the frictionless standard fields (columns) is extended as described below. 
Note that each field must have a unique name; otherwise, they can not be unambiguously addressed. If several columns share the same name, the names must be converted first.
The units should preferably be represented by a string that can be interpreted by a program used in the following data evaluation process. Astropy's notation is leveraged for unitpackage.\cite{ASTROPY2013,ASTROPY2022}

\subsubsection{Exploring Data}

Once local files are annotated, browsing and filtering by metadata is facilitated for programmatic approaches. For example, dedicated methods allow loading all metadata files in a directory and filtering them for specific keys or values, such as a user or a sample. This aspect is elaborated on in the following section.

\subsection{Data Packages}

The files created in the previous section (CSV and YAML) are now transformed into a frictionless Data Package. 
A simple Data Package consists of a single \mintinline{text}{resource} in a JSON (\mintinline{text}{package.json}). The structure of the file-based resource \mintinline{text}{data.csv} is described in the \mintinline{text}{schema} key. Preferably, the resource contains additional information on the fields (such as units and descriptions) and other metadata from the YAML file describing the data. 

We create Data Packages containing a single resource, denoted as unitpackage. To interact and create such unitpackages we created the \mintinline{text}{unitpackage} API\cite{UNITPACKAGE2024}, which is based on the frictionless framework. Using the YAML and CSV from Fig.~\ref{fig:csv_metadata}, a Data Package can be created as follows.

\begin{minted}[
frame=lines,
framesep=1mm,
baselinestretch=1.0,
% bgcolor=LightGray,
breaklines=true,
fontsize=\scriptsize,
linenos
]{python}
from unitpackage.entry import Entry
import yaml

with open("./data/raw/data.csv.meta.yaml", "rb") as f:
    metadata = yaml.load(f, Loader=yaml.SafeLoader)

fields = metadata["figure description"]["fields"]

entry = Entry.from_csv(csvname="/data/raw/data.csv", metadata=metadata, fields=fields)
entry.save(outdir="./data/generated/")
\end{minted}

The code will store the CSV and JSON in the specified \mintinline{text}{outdir}. The resulting JSON for the data provided in Fig.~\ref{fig:csv_metadata} is available in the supporting information (including the code below).

A collection of unitpackages can be loaded with the \mintinline{text}{unitpackage} API to browse, explore, modify or visualize the entries. The \mintinline{text}{unitpackage} documentation provides a detailed description and usage examples. Excerpted are some relevant features. A collection of unitpackages can be loaded from a specific directory, such as the \mintinline{text}{outdir} used above.

\begin{minted}[
frame=lines,
framesep=1mm,
baselinestretch=1.0,
% bgcolor=LightGray,
fontsize=\scriptsize,
breaklines=true,
linenos
]{python}
>>> from unitpackage.collection import Collection
>>> 
>>> db = Collection.from_local('./data/generated')
>>> db
[Entry('data')]
\end{minted}

The collection can be filtered to return a new collection based on a predicate in the metadata or boundary conditions defined on the underlying data. Alternatively, a single entry can be selected by its identifier (lowercase data filename without suffix). The descriptors defined in the YAML are accessible through that entry. 

\begin{minted}[
frame=lines,
framesep=1mm,
baselinestretch=1.0,
% bgcolor=LightGray,
breaklines=true,
fontsize=\scriptsize,
linenos
]{python}
>>> entry = db['data']
>>> entry.user
Max Doe
\end{minted}

The data from the CSV can be loaded as a \mintinline{text}{pandas} DataFrame, one of the very prominent frameworks to work with tabular data.\cite{PANDAS2010,PANDAS2020software}

\begin{minted}[
frame=lines,
framesep=1mm,
baselinestretch=1.0,
% bgcolor=LightGray,
breaklines=true,
fontsize=\scriptsize,
linenos
]{python}
>>> entry.df
   t     U    T
0  0  1.01  275
1  1  1.02  275
2  2  1.05  275
...
\end{minted}

First simple operations with units are implemented.

\begin{minted}[
frame=lines,
framesep=1mm,
baselinestretch=1.0,
% bgcolor=LightGray,
breaklines=true,
fontsize=\scriptsize,
linenos
]{python}
>>> entry.rescale({'U':'V'}).df
   t        U    T
0  0  0.00101  275
1  1  0.00102  275
2  2  0.00105  275
...
\end{minted}

The metadata and data can now be used in conjunction, for example, to calculate new quantities. In the example data in Fig.~\ref{fig:csv_metadata}, the resistance of a resistor should be determined when a current is applied to it, and the voltage across the resistor is measured. The resistance can be calculated using Ohm's law ($R = U/I$).

\begin{minted}[
frame=lines,
framesep=1mm,
baselinestretch=1.0,
% bgcolor=LightGray,
fontsize=\scriptsize,
breaklines=true,
linenos
]{python}
>>> import astropy.units as u
>>> 
>>> # Value from metadata
>>> I = entry.current
>>> # Mean voltage from the actual data
>>> U = entry.df['U'].mean() * u.Unit(entry.field_unit('U'))
>>> R = U / I
>>> R.to(u.Ohm)
0.2 Ohm
\end{minted}

Here, \mintinline{python}{I} is an astropy quantity. The values in the \mintinline{text}{pandas} DataFrame are dimensionless, but the units can be retrieved from the field description to construct an \mintinline{text}{astropy} quantity, which allows computing the resistance \mintinline{python}{R}, also as an \mintinline{text}{astropy} quantity object.

The supporting information provides further examples, illustrating how (i) databases of unitpackages can be filtered for specific properties, (ii)  unitpackages can be created from existing data or pandas DataFrames, (iii) collections can be retrieved from remote repositories, containing for example, supporting data to original works\cite{ENGSTFELD2024RuData,ENGSTFELD2024Ru} and (iv) specific classes of entries and databases can be created, which provide for example, direct access to the certain properties such as \mintinline{python}{entry.R}.

\subsubsection{Standardizing Raw Data}

The aforementioned examples showing the usage of the \mintinline{text}{unitpackage} API are designed to work with CSV, containing a single header line above the lines containing the comma-separated values (according to the W3C recommendation for tabular data%CSV standard
). The structure of output data recorded with existing in-house software or software from a proprietary device usually deviates from such an ideal CSV structure. The files usually contain several header lines and/or footer lines, which can vary between measurements with different routines and might as well come in different formats. 

In such cases, we suggest writing a loader that creates a simple \mintinline{text}{pandas} DataFrame, with a single header line and using the approach to store the \mintinline{text}{pandas} DataFrames as Data Package with the \mintinline{text}{unitpackage} API (see supporting information). The metadata from the header and footer can be (processed and) included in the metadata of the unitpackage.

Another issue is that the column names for files recorded with different software (controlling different devices) are different for the same type of measurement. For example, the header name for a recorded voltage could be given as \mintinline{text}{U / V}, \mintinline{text}{voltage [mV]}, or \mintinline{text}{volt. [W / A]}. In that case, it is useful to define an internal standard, such that all files are stored simply using, e.g., \mintinline{text}{U}, for the measured voltage. As illustrated above, storing the units in the header name is unnecessary since they are included in the Data Package resource. Note that only the structure changed in the converted files, and the data is still identical to that in the original files.

\subsection{Demonstrators}
\label{sec:demonstrators}

The advantages and use cases of storing data and making metadata machine-readable are illustrated using unitpackages for electrochemistry data, with tools developed by the echemdb community.\cite{ECHEMDB}
The data pertain to studies on the electrochemical properties of well-defined single-crystal electrodes studied by a commonly employed technique (cyclic voltammetry).
The materials used are very well-defined, and the measurement technique is well-established within the community. 
Over recent decades, such studies have provided significant fundamental insights, aiding in the understanding of more complex three-dimensional systems found in applied research and practical applications. 
This makes it an ideal benchmark for comparing experimental data, published figures, and computational results.

\subsubsection{Electrochemistry Metadata Schema}

To date, schemas to describe electrochemical systems fulfilling our requirements are unavailable. There are, however, a few resources and ontologies describing electrochemical terms in general.\cite{BARD2012,PINGARRON2020}
Also, terms used to describe battery-related systems might be helpful.\cite{CLARK2022,STIER2024}
Linking the terms of our metadata schema to these ontologies, for example, using JSON-LD\cite{JSONLD} is envisioned.
This aspect is, however, beyond the scope of this work. For further information the reader is referred to other projects, which use such an approach.\cite{ROCCA2019,JACOB2020,BLUMBERG2021}

Due to the lack of a metadata schema, a schema for electrochemical data was developed using the approach with YAML templates described above.\cite{ECMDS}
The descriptors were defined based on the systems investigated in our laboratory and those reported by other groups. By including data from other research groups, we could identify descriptors that might not have been relevant to our work.
Over time, the metadata schemas have been adapted, simplified, and restructured to create generalised objects common for any research area (curation, projects, figure description, and experimental) and system-specific objects, which include our information for electrochemical data.
In a final step, a JSON schema was developed,\cite{ECMDS} which is used to validate the YAML metadata for actual data. An extensive YAML file can be found in the supporting information and Ref.~\citenum{ECMDS}.

\begin{figure*}[!ht] %Add * to span figure over two columns
    \centering
    \includegraphics[width=1\textwidth]{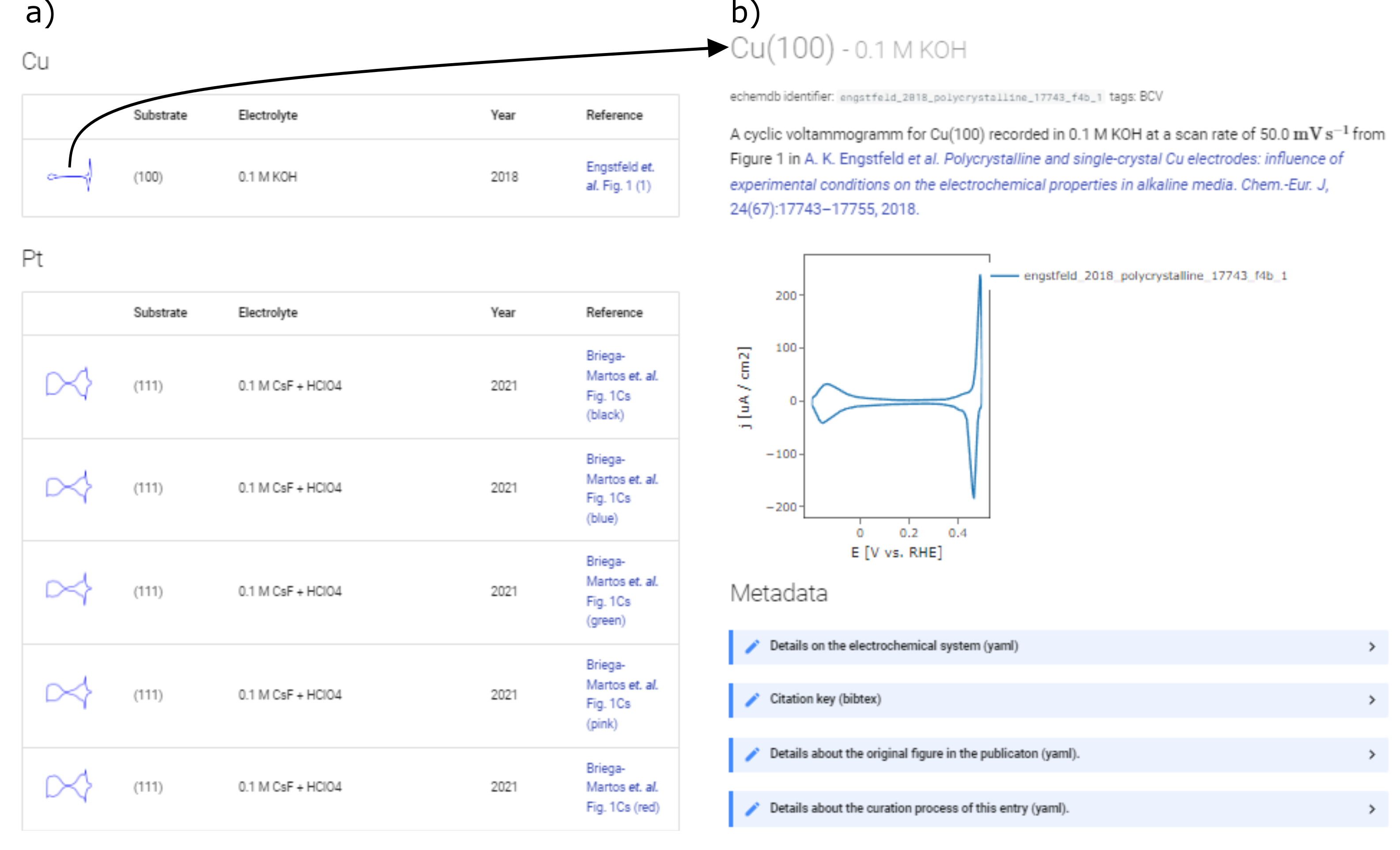}
    \caption{Snapshot of the CV database displaced on the echemdb website\cite{ECHEMDBwebsite} generated from frictionless Data Packages. Page a) shows a list of entries with the most relevant descriptor and b) provides detailed information on the respective entry. }
    \label{fig:website}
\end{figure*}

\subsubsection{Literature Data}

Scientific results are often compared with and validated against data in scientific publications, but such data is usually only available as a plot in a printed (PDF) version. 
We developed “svgdigitizer”, which extracts the data from curves carefully retraced using scalable vector graphics (SVG)
.\cite{SVGDIGITIZER2023}
An outstanding feature of svgdigitizer is that the workflow based on SVG allows for a relatively easy review process of the digitized data with having the original PDF file as only prerequisite.
In combination with a YAML file containing the metadata extracted manually from the publication (following the schema for electrochemical data described above), a unitpackage can be created.
We collected and stored such YAML and SVG files for electrochemical data in a git repository, which validates the YAML files against a JSON schema\cite{ECEHMDATA} and creates unitpackages from all files upon merging pull requests. The output packages are created for specific versions of the metadata schema.

\subsubsection{Specific Unitpackage Module}

To interact and explore the cyclic voltammetry data, we added \mintinline{python}{CVEntry} and \mintinline{python}{CVCollection} to the \mintinline{text}{unitpackage} module, with which certain descriptors in the metadata can be accessed more conveniently, or statistics on the collection can be displayed. Also, standard plots of type $I$--$E$ or $j$--$E$ (or $U$ instead of $E$) for displaying the CVs can be created. The data can be collected from the GitHub repository using the \mintinline{text}{unitpackage} API. Storing electrochemical data locally using the same metadata structure allows for directly comparing measured and published data.

\begin{minted}[
frame=lines,
framesep=1mm,
baselinestretch=1.0,
breaklines=true,
fontsize=\scriptsize,
linenos
]{python}
>>> from unitpackage.cv.cv_collection import CVCollection
>>> 
>>> cvdb = CVCollection.from_remote()
>>> cvdb.describe()
{'number of references': 44,
 'number of entries': 205,
 'materials': {'Ag', 'Au', 'Cu', 'Pt', 'Ru'}}
\end{minted}

\subsubsection{Displaying and Organizing Collections}

The machine-readable experimental or literature collections can, in combination with the \mintinline{text}{unitpackage} API, be displayed in an appealing way using web frameworks / technologies for creating static or dynamic websites. The frictionless toolbox offers Livemark for this purpose. To display the electrochemical collection on a website,\cite{ECHEMDBwebsite} we used MKDOCS.\cite{MKDOCS} 
On the website, the data is sorted in overview pages based on the material used for the electrodes, as illustrated in a snapshot in Fig.~\ref{fig:website}a. The table contains a thumbnail of the measured curve generated from the CSV and descriptors commonly used to describe the electrochemical system. For each entry, a separate page is generated as illustrated for a single entry in Fig.~\ref{fig:website}b. The page provides information on the entrys' source, the most important descriptors to the community, an interactive plot, and collapsible tabs for the different metadata categories available, extracted and generated from the unitpackage.

Most simply, one can explore, organize, evaluate, and describe the data in Jupyter notebooks. Such notebooks can also be served using web frameworks. One rather straightforward solution is the use of Jupyter Book.\cite{JupyterBook} This interface was used to document the workflow described in the first two sections of this work.\cite{rawtofigure} This approach is an optimal solution to create an interactive ELN locally from files in the file system, which can be made available along with a scientific publication.

%%%%%%%%%%%%%%%%%%%%%%%%%%%%%%%%%%%%%%%%%%%%%%%%%%%%%%%%%%%%%%%%
%%%%%%%%%%%%%%%%%%% --- Conclusion --- %%%%%%%%%%%%%%%%%%%%%%%%%
%%%%%%%%%%%%%%%%%%%%%%%%%%%%%%%%%%%%%%%%%%%%%%%%%%%%%%%%%%%%%%%%
\section{Conclusion}
 
In this work, we showcase a lightweight and time-saving approach to automatically annotating data with machine-readable metadata. Converting the annotated files into frictionless Data Packages improves their interoperability in general. The unitpackage API allows exploring and visualising the data. Furthermore, it provides a simple, extensible framework for evaluating data collections, making use of both the values from the actual data and the information in the metadata. 
We are confident that the approach presented here will help raise awareness of structured data and metadata in institutions with little knowledge of this topic or are still undecided about which high-level solution should be introduced. The approach might also interest those who want to share their data in an appealing way.

%%%%%%%%%%%%%%%%%%%%%%%%%%%%%%%%%%%%%%%%%%%%%%%%%%%%%%%%%%%%%%%%
%%%%%%%%%%%%%%%%%%% ---Example Section - delete later --- %%%%%%%%%%%%%%%%%%%%%%%%%
%%%%%%%%%%%%%%%%%%%%%%%%%%%%%%%%%%%%%%%%%%%%%%%%%%%%%%%%%%%%%%%%


\section{Data Availability}
The data supporting the findings of this study are openly available in Zenodo at \url{https://zenodo.org/doi/10.5281/zenodo.13739617} or directly from the GitHub repository at \url{https://github.com/echemdb/manuscript_echemdb_rdm/}. 

\begin{acknowledgement}

AKE gratefully acknowledges support by the DFG (German Science Foundation) through the collaborative research centre
SFB-1316 (project ID: 327886311). We thank all curators that provided literature data for the echemdb website.

\end{acknowledgement}

%\begin{suppinfo}
%
%``Experimental procedures and
%characterisation data for all new compounds. The class will
%automatically add a sentence pointing to the information on-line:
%
%\end{suppinfo}

\bibliography{bibliography.bib}

\begin{mcitethebibliography}{39}
\bibitem[SUF()]{SUFFIX}
Including the suffix of the data file in the metadata file is mostly relevant
  when a data acquisition software stores different data types with the same
  name but different suffixes. An example would be where the acquisition
  software stores along with the \mintinline{text}{data.csv} a
  \mintinline{text}{data.yaml}, which contains information, i.e., on the device
  used to perform a measurement.\relax
\mciteBstWouldAddEndPunctfalse
\mciteSetBstMidEndSepPunct{\mcitedefaultmidpunct}
{}{\mcitedefaultseppunct}\relax
\EndOfBibitem
\end{mcitethebibliography}

\end{document}